\documentclass[aps,twocolumn,prb,superscriptaddress,floatfix]{revtex4}
\usepackage{amsmath,amssymb,graphicx}
\usepackage{hyperref,epsfig,subfigure}
\renewcommand{\vec}[1]{\mathbf{#1}}
\newcommand{\sech}{\mathop{\mathrm{sech}}}

\begin{document}

\newcommand{\paule}[1]{$\spadesuit${\sl #1}}
\newcommand{\tgunkv}[1]{{\vec{#1},t}}
\newcommand{\tgunk}{\tgunkv{k}}
\title{Finite momentum condensation in a pumped microcavity
}

\author{R. T.  Brierley}\affiliation{Cavendish Laboratory, University
  of Cambridge, Cambridge CB3 0HE, United Kingdom.}
\author{P. R. Eastham}\affiliation{Cavendish Laboratory, University of Cambridge, Cambridge CB3 0HE, United Kingdom.}
\affiliation{School of
  Physics, Trinity College, Dublin 2, Ireland.}

\begin{abstract}
  We calculate the absorption spectra of a semiconductor microcavity
  into which a non-equilibrium exciton population has been pumped. We
  predict strong peaks in the spectrum corresponding to collective
  modes analogous to the Cooper modes in superconductors and fermionic
  atomic gases. These modes can become unstable, leading to the
  formation of off-equilibrium quantum condensates. We calculate a
  phase diagram for condensation, and show that the dominant
  instabilities can be at a finite momentum. Thus we predict the
  formation of inhomogeneous condensates, similar to
  Fulde-Ferrel-Larkin-Ovchinnikov states.\end{abstract}
  
\maketitle

\section{Introduction}
\label{sec:introduction}

The appearance of order at an equilibrium phase transition is a
central concept in many areas of physics, from condensed matter to the
physics of the early universe. Recently there has been considerable
interest in the more general problem of ordering far from thermal
equilibrium, motivated by the possibility of quantum quench
experiments in cold atomic gases.\cite{bloch_many-body_2008} In a
quench the parameters of the system are rapidly switched from a
disordered to an ordered phase, and the disordered state forms the
initial conditions for a dynamics with the new parameters. An
interesting regime is that of coherent relaxationless dynamics, which
can lead to the formation of non-equilibrium order including
crystallization,\cite{kollath_quench_2007} condensation, and
ferromagnetism.\cite{babadi_non-equilibrium_2009}

Among condensed matter systems, semiconductor microcavities are
promising candidates for studying such quench dynamics. The
non-equilibrium dynamics of microcavities has attracted considerable
interest both experimentally
\cite{richard_spontaneous_2005-1,lai_coherent_2007,kasprzak_formation_2008,maragkou_spontaneous_2009,sanvitto_exciton-polariton_2009,krizhanovskii_coexisting_2009}
and theoretically,
\cite{szymanska_mean-field_2007,wouters_excitations_2007,keeling_spontaneous_2008,eastham_mode_2008,doan_condensation_2005}
with recent experiments demonstrating regimes where the low-energy
quasiparticles, polaritons, form a condensate. More recently, an
experiment has been proposed to implement a quantum
quench,\cite{eastham_quantum_2007} by rapidly preparing a microcavity
in a non-condensed initial state. The coherent dynamics of this
non-condensed state is predicted to lead to a form of non-equilibrium
condensation, similar to that predicted in a quenched Fermi
gas.\cite{andreev_nonequilibrium_2004,barankov_atom-molecule_2004,yuzbashyan_relaxation_2006}
We show here that, as in the Fermi gas, such condensation is due to
the appearance of a new collective mode. Moreover, we show that in the
microcavity the dominant instability occurs at a finite wavevector.
Thus we predict that microcavities could be used to realize
inhomogeneous condensates,\cite{casalbuoni_inhomogeneous_2004} i.e.,
those characterized by a spatially varying phase. These condensates
are similar, in this essential respect, to those predicted by Fulde,
Ferrel, Larkin and Ovchinnikov (FFLO) in unbalanced Fermi systems.
\cite{casalbuoni_inhomogeneous_2004}

In this paper, we first calculate the optical spectra of a microcavity
a short time after it has been prepared in a non-condensed state,
i.e., immediately after the ``quench''. We find that the collective
mode responsible for condensation is directly observable in these
spectra. We use this analysis to calculate a phase diagram for the
non-equilibrium condensation, and show that the condensation generally
occurs at a finite momentum.  While we focus on a microcavity
containing quantum dots, our analysis is based on the Maxwell-Bloch
equations. These describe a wide variety of coupled light-matter
systems, implying a broad relevance of our work.

The remainder of this paper is structured as follows. In
Sec.~\ref{sec:model} we briefly review the proposed quench experiment,
and outline our model. In Sec.~\ref{sec:absorption-spectra} we present
absorption spectra of the system. In Sec.~\ref{sec:phase-diagram} we
discuss the phase diagram and the possibility of finite momentum
condensation, and in Sec.~\ref{sec:discussion} we discuss the
connections to FFLO and the role of nonlinear terms. Section\
\ref{sec:conclusion} summarizes our conclusions. Finally, the appendix
contains a brief treatment of the preparation of non-condensed initial
states by optical pumping.

\section{Model}
\label{sec:model}

We consider an experiment, proposed in
Ref. \onlinecite{eastham_quantum_2007}, on a set of localized exciton
states in a planar semiconductor microcavity. Such excitonic states
could be realized in practice using either highly disordered quantum
wells (where excitons are localized by disorder) or quantum dots. The
proposed experiment involves two stages which are separated in time
and can be regarded as independent. In the first stage, the localized
states are driven by a chirped laser pulse. This pulse creates an
energy-dependent population in the inhomogeneously broadened exciton
line by adiabatic rapid passage (ARP). For certain populations a
second stage may then occur, where the population evolves into a
non-equilibrium condensate due to the photon-mediated interactions
between the excitons.

As in Ref.\ \onlinecite{eastham_quantum_2007} we describe the system
using a generalization of the Dicke model.\cite{keeling_bcs-bec_2005} The localized exciton states are treated
as two-level systems, with the standard dipole coupling to the
electromagnetic field. The state localized at site $i$ is specified by
the Bloch vector $\vec\sigma_i=\langle \hat \sigma_i \rangle$, where
the inversion $\sigma^z_i=1 (-1)$ for an occupied (unoccupied) state,
and $\hat \sigma_i^-$ is the exciton annihilation operator. Angle
brackets $\langle\rangle$ denote expectation values in the quantum
state of the system.

We consider time scales short compared with the exciton lifetime,
which is at least $100\; \mathrm{ps}$,\cite{langbein_microscopic_2005}
and treat the electromagnetic field using a mean-field
approximation. In this approximation the photon creation and
annihilation operators are replaced with their expectation values, and
hence become c-numbers. The resulting equations of motion are linear
in the remaining operators, so that we may take their expectation
values without further approximation. The resulting dynamics obeys the
generalized Maxwell-Bloch equations:
\begin{gather}
 \label{eq:eofm1}
 i\dot\psi_{\vec{k}}=\omega_{\vec{k}} \psi_\vec{k} + g \int
 P_\vec{k}\,dE+f_\vec{k}+F\delta_{\vec{k-p}}\text{,}\\
 \label{eq:eofm2}
 i\dot
 P_\vec{k}=EP_\vec{k}-g\sum_\vec{k'}D_\vec{k-k'}\psi_\vec{k'}\text{,}\\
 \label{eq:eofm3}
 i\dot D_\vec{k} = 2g\sum_\vec{k'} \left(P^*_\vec{k'-k}
     \psi_\vec{k'}-P_\vec{k'+k} \psi^*_\vec{k'}\right)\text{.}
\end{gather}
Here $\psi_\vec{k}$ is the complex amplitude of the microcavity mode
with in-plane wavevector $\vec k$ and energy $\omega_{\vec{k}}$
($\hbar=1$). It is related to the expectation value of the photon
annihilation operator by $\psi_\vec{k}=\langle\hat
\psi_\vec{k}\rangle/\sqrt{N}$, where $N$ is the total number of
localized states. This normalization is convenient when dealing with
condensation, since macroscopic occupation corresponds to a finite
$\psi_\vec{k}$ in the thermodynamic limit $N\to\infty$. We allow for
the finite lifetime of the photon modes by taking $\Im \omega_{\vec
  k}=-\gamma$. $f_\vec{k}$ is introduced to allow us to calculate the
linear response. $F$ is an externally applied pump field, with
wavevector $\vec{p}$, that is used to create the non-equilibrium
population.

The coupling $g=g_i\sqrt{n}$ in Eqs. (\ref{eq:eofm1}-\ref{eq:eofm3})
is related to the dipole coupling strengths of the localized states,
$g_i$, and their area density $n$.  To simplify the notation we have
taken $g_i$ to be the same for all states; the extension to a
distribution is straightforward. In the dipole gauge \begin{equation}
  g_i=d \sqrt{\frac{E_i}{2\epsilon_0\epsilon w}},\end{equation} where
$d$ is the matrix element of the dipole operator $e\hat r$ between the
zero-exciton and one-exciton states, and $E_i$ their energy
difference. $w$ is the effective width of the cavity, which arises
from the normalization of the cavity mode functions.

$P_{\vec{k}}(E)$ is the collective polarization of the ensemble at
wavevector $\vec k$ due to states with energy in a small interval near
$E$,
\begin{equation}
 \label{eq:collpol}
 P_\vec{k}(E)\delta E = \frac{1}{N}\sum_i\phantom{}^{'}
 \left<\hat\sigma_i^-\right> e^{-i\vec{k}\cdot{\vec{r}_i}}\text{,}
\end{equation}
where the prime indicates that the sum runs over states with exciton
energies between $E$ and $\delta E$.  $D_\vec{k}(E)$ is the collective
inversion, defined in a similar way with $\hat\sigma^z$ replacing
$\hat\sigma^-$.

In the following we shall be concerned with large $N$, and wavevectors
which are small compared with the inverse spacing of the localized
states. In these limits we may approximate sums over dot positions,
such as those in Eq. (\ref{eq:collpol}), by
\begin{equation}
  \label{eq:disorderapprox}
  \frac{1}{N}\sum_i\phantom{}^{'}e^{-i\vec k \cdot \vec r_i}\approx\nu(E)\delta_{\vec
    k,0}\delta E+O\left(\frac{1}{\sqrt{N}}\right).
\end{equation} Here $\nu(E)$ is the distribution of
localized states in energy, normalized to one. Thus $N\nu(E)\delta E$ is the
number of terms in the primed sum, Eq. (\ref{eq:disorderapprox}). For $\vec k\neq0$ the phasor sum is a 2D random walk, producing the $O(1/\sqrt{N})$ corrections.\cite{goodman_statistical_1975}

The approximation of Eq. (\ref{eq:disorderapprox}) corresponds to
replacing the response of the disordered dielectric with its
homogeneous average response, so that the wavevector is
well-defined. This is similar to the linear dispersion model which has
been extensively used for inorganic
microcavities.\cite{savona_effect_2007} The $O(1/\sqrt{N})$
corrections describe Rayleigh scattering from density fluctuations in
the dielectric. They are generally small corrections because the
scatterers are dense, so that on long wavelengths the medium appears
homogeneous. The corrections can become important at very small or
large wavevectors,\cite{agranovich_cavity_2003,litinskaya_loss_2006}
for modes whose group velocity becomes very small. In this case a long
lifetime is required for the wavevector to be well-defined, so that
even weak scattering or absorption destroys the quasi-propagating
modes. Here, however, we are concerned with modes that have a
significant dispersion due to their photon component. Furthermore, the
lifetime of these modes is massively enhanced by resonant gain from
the populated excitons. Thus the leading approximation of
Eq. (\ref{eq:disorderapprox}) will capture the physics at the
experimentally relevant wavevectors.

\section{Absorption spectra}
\label{sec:absorption-spectra}

In the experiment $F$ is a chirped pulse, which creates a
non-equilibrium population of excitons using ARP. In the appendix we
demonstrate this explicitly, using a model pulse,
Eq.~\eqref{eq:pulse-params}, for which analytical solutions to the
dynamics exist. Following the pulse the exciton states are populated
with a distribution given by Eq.~\eqref{eq:popdist} and the fields and
polarizations are negligible $\psi_\vec{k}\approx 0$,
$P_\vec{k}\approx 0$.

To establish the optical properties of the microcavity immediately
after the pump pulse we find the response to a weak probe
$f_\vec{k}$. The susceptibility can then be found from the induced
electromagnetic field $\delta
\psi_\vec{k}\equiv\sum_{\vec{k'}}\int\chi_\vec{k,k'}(t-t')f_\vec{k'}(t')\,dt'$. If
the system is stable then $\psi_\vec{k}$ and $P_\vec{k}$ are small (of
order $f_{\vec{k}}$) for all times, whereas if it is unstable they are
only small soon after the pump pulse. In both regimes we may neglect
terms above first order in $\psi_\vec{k}$ and
$P_\vec{k}$. Eq.~\eqref{eq:eofm3} then gives $\dot D_\vec{k}=0$, so
the non-equilibrium population is constant.  Fourier transforming the
linearized Eqs. \eqref{eq:eofm1} and \eqref{eq:eofm2} gives
\begin{equation}
 \label{eq:responpsistep}
 \omega\delta\psi_\vec{k}=\omega_\vec{k}\delta\psi_\vec{k}-g^2\sum_\vec{k'}\int\frac{D_{\vec{k-k'}}\delta\psi_\vec{k'}}{\omega-E}dE+f_\vec{k}\text{.}
\end{equation}

The pumping populates the states independently of their position,
within the pump spot. Thus the sum in $D_\vec{k-k'}$ is strongly
peaked near the forward scattering direction $\vec{k-k'}=0$, as
discussed above [Eq. (\ref{eq:disorderapprox})]. Neglecting the
smaller off-diagonal scattering terms we obtain a diagonal response
function
\begin{equation}
 \label{eq:responsefunc}
  \chi_\vec{k}(\omega)=\frac{1}{\omega-\omega_\vec{k}+g^2\int
   \frac{D_0(E)}{\omega-E}\,dE}\text{.}
\end{equation} 

The absorption coefficient of the microcavity follows from the
susceptibility\cite{mahan_many-particle_1990,eastham_bose_2001}
\begin{equation}
  A(\omega)=-2\lim_{\epsilon\to0}\mathrm{Im}
  \chi(\omega+i\epsilon),\end{equation} where the infinitesimal
$\epsilon$ appears due to causality, and can be physically understood
as a small damping constant for the excitons. The sign is such that $A(\omega)>0$ corresponds to absorption of energy by the system. Thus from
Eq.~\eqref{eq:responsefunc} we obtain \begin{eqnarray} A(\omega)&=&
  2\frac{\gamma-g^2\pi D_0(\omega)}{\left(\omega-\omega_c +
      g^2\mathcal{P}\int\frac{
        D_0}{\omega-E}dE\right)^2+\left(\gamma-g^2\pi
      D_0(\omega)\right)^2} \nonumber \\ &=&
  2\frac{H(\omega)}{G(\omega)^2+H(\omega)^2}.  \label{eq:absspec}
\end{eqnarray} When the dots are unoccupied $D_0(\omega)<0$ and the empty exciton
states contribute to absorption. For energies where there are occupied
exciton states $D_0(\omega)>0$, describing gain due to the
population. 

In general the response, Eq. \eqref{eq:absspec}, peaks near the zeroes
of $G(\omega)$, which are at the energies of the normal modes. These
energies differ from the energy of the cavity resonance due to the coupling
to the exciton states. For an unpopulated state the condition
$G(\omega)=0$ recovers the usual polariton energies of the Lorentz
oscillator model,\cite{burnsssp} but in general the spectrum differs
due to the presence of the non-equilibrium population. The modes have
a lifetime determined by the second factor in the denominator, with
contributions from the cavity losses and the resonant mixing with the
band of exciton states.  As expected it is damping which controls the
overall strength of the absorption, so that the damping factor
$H(\omega)$ also appears in the numerator.

The only dependence of the spectra, Eqs.~\eqref{eq:responsefunc} and
\eqref{eq:absspec}, on the wavevector is in the energy of the cavity
mode, $\omega_{c}=\Re(\omega_{\vec k})\approx \omega_{0} +
|\vec{k}|^2/(2m)$. We therefore show results as functions of $\omega_{c}$,
which corresponds experimentally to both the incident probe angle and
the cavity width.

\begin{figure}[t]  
 \centering \subfigure[Populated]{
   \includegraphics[width=8.6cm]{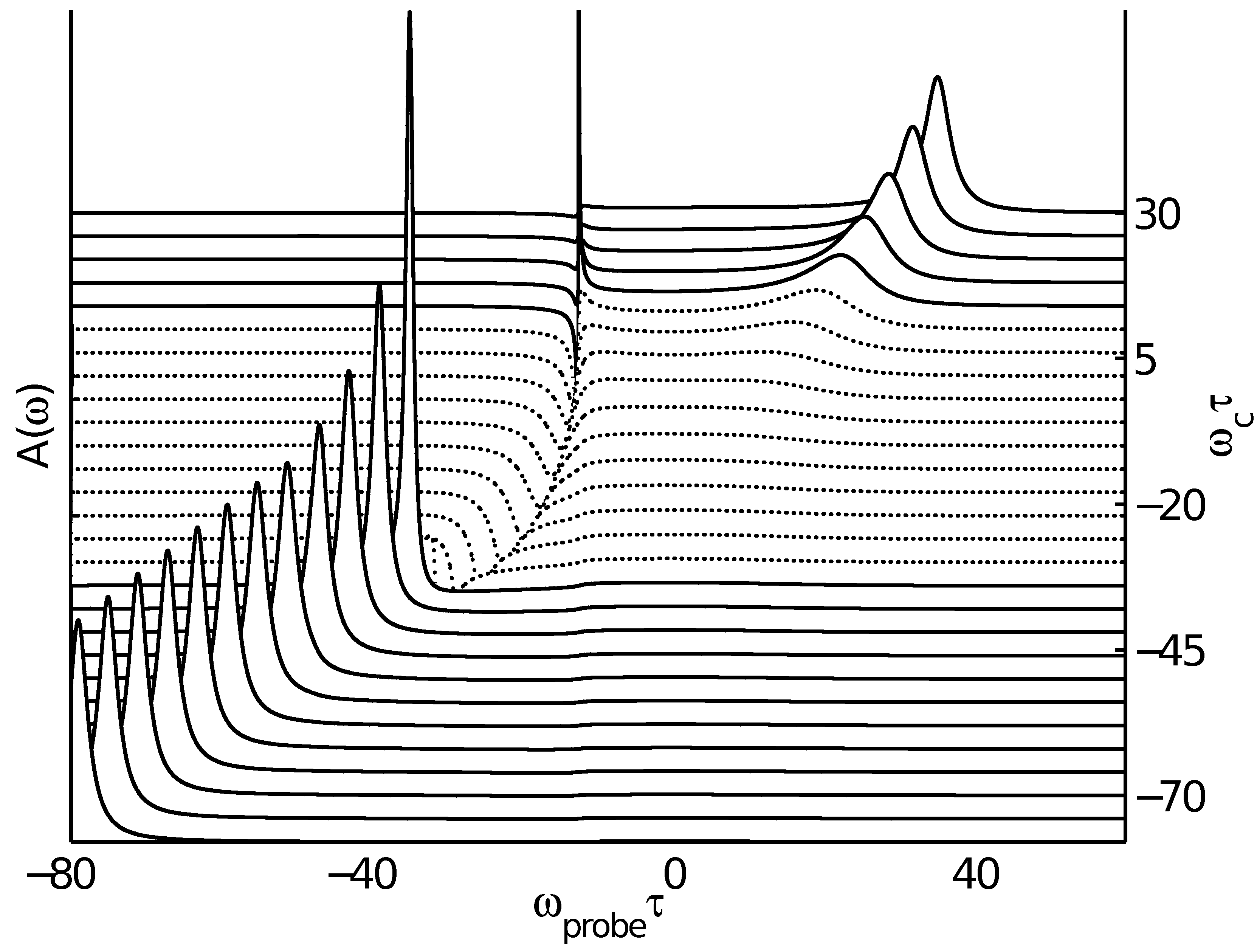}\label{fig:popabsspec}}

 \subfigure[Unpopulated]{
   \includegraphics[width=8.6cm]{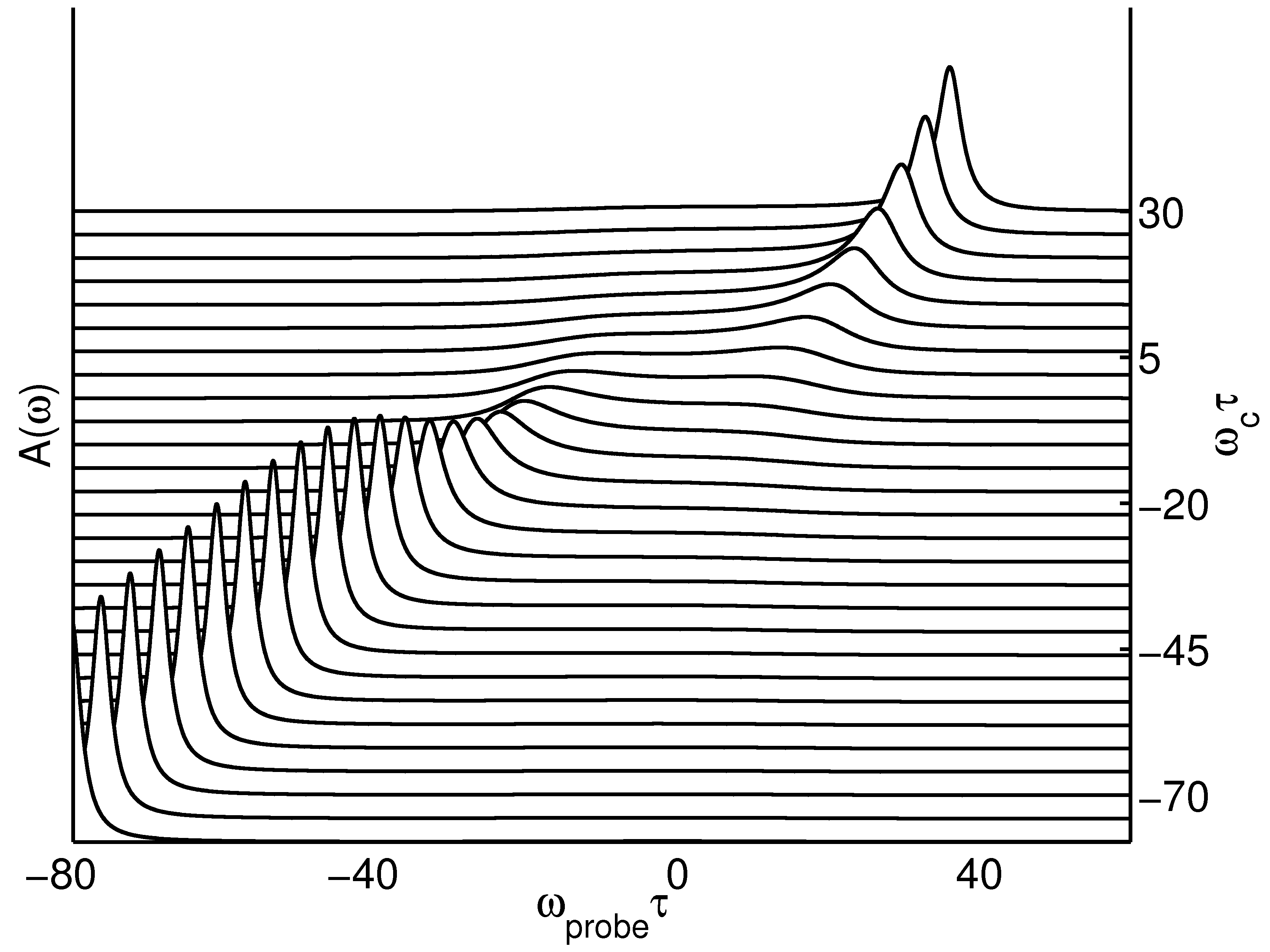}\label{fig:vacabsspec}}
 \caption{Absorption spectra [equation \eqref{eq:absspec}] in
   arbitrary units, for the parameters described in the text. Top
   panel: spectra with the exciton population created by the pump
   pulse. Bottom panel: spectra of the unpopulated microcavity. Each
   curve corresponds to a different value of the cavity mode energy
   $\omega_c$, vertically offset as indicated by the right-hand
   axes. Note the peak developing in the populated system at
   $\omega_\text{probe}\approx-13/\tau$, indicating the presence of a
   collective mode. There are regions of probe gain ($A(\omega)<0$)
   between $-31<\omega_{probe}\tau<-13$, where gain from the populated
   excitons overcomes the cavity losses. Dotted lines indicate spectra
   with unstable normal modes. Instabilities occur when a normal mode
   lies in the region of gain and in these cases the absorption
   spectra have negative peaks, which are hidden from view in this
   figure.}
 \label{fig:absspec} 
\end{figure}

Figure~\ref{fig:absspec} illustrates absorption spectra obtained from
Eq.~\eqref{eq:absspec} for both a pumped and unpumped exciton
line. These spectra are valid at all times if condensation does not
occur (see later), but only soon after the pump pulse if it does. We
have taken a Gaussian model for the inhomogeneously-broadened exciton
line, with standard deviation $\sigma$, and measure energies relative
to the center of the line. We choose the duration $\tau$ of the pump
pulse as our unit of energy, and have taken $g=13/\tau,
\gamma=1.5/\tau, \sigma=15/\tau$. These parameters, with
$\tau=3\;\mathrm{ps}$, are reasonable for a microcavity containing
strongly disordered quantum wells.~\cite{piper_conf} As discussed in the appendix the
pump creates a population equivalent to a Fermi function with
temperature $1/(\pi k_B\tau)$, and Fermi energy $\mu$ dictated by the
chirp and center frequency of the pulse; we choose a pulse for which
$\mu=-12.5 /\tau$.

The lower panel of Fig.\ \ref{fig:absspec} shows the expected result
for an unpopulated microcavity. There is a pronounced peak in the
absorption at the cavity mode energy, which broadens as the cavity
mode is tuned through the excitons. There is some suggestion of an
anticrossing near resonance, i.e., a polariton splitting, but since the
inhomogeneous broadening is relatively large compared with the coupling
this is a weak effect. The top panel shows that the population
dramatically changes the absorption spectrum. For these parameters it
leads to a range of probe frequencies for which $\gamma<g^2\pi
D_0(\omega)$, and the absorption coefficient, Eq.~\eqref{eq:absspec},
becomes negative. This occurs when the gain from the populated exciton
states exceeds the losses, so that there is a net gain for the probe
beam. Moreover, we see a pronounced additional peak in the absorption
spectrum, which first appears near the upper edge of the population as
the cavity mode energy is decreased. As the cavity energy is decreased
still further this peak moves down through the region of gain, before
the spectrum finally reverts to one dominated by the unperturbed
cavity mode.

This additional peak in the absorption spectrum is analogous to the
Cooper pairing mode in a superconductor or Fermi gas, that gives rise
to the Cooper instability. The analogy can be seen by noting that the
normal-mode condition $G(\omega)=0$ contained in
Eq.~\eqref{eq:absspec} is the Cooper equation, as discussed for this
system in Ref.\ \onlinecite{eastham_quantum_2007}.  The
non-equilibrium exciton population corresponds to the Fermi
distribution, while the photon-mediated interaction between excitons
corresponds to the pairing interaction between the electrons. As in a
superconductor the sharp step in the population leads to collective
modes generated by the pairing interaction. Fig.\ \ref{fig:absspec}
shows that, for reasonable parameters, these collective modes give
rise to strong features in the spectra.

It is interesting to compare the spectra of Fig.\ \ref{fig:absspec}
with the predictions for an equilibrium condensate in the same
model.\cite{eastham_bose_2001} In that case the condensation opens a
gap in the single particle spectrum, which is the analog of the Cooper
gap of the superconductor. Inside this gap is a collective mode, which
is the analog of the Cooper mode or phase mode of the superconductor.
The features visible in Fig.\ \ref{fig:absspec} arise from the
non-equilibrium generalization of the collective mode (which is a
different spectral feature than the gap). It is clear from Fig.\
\ref{fig:absspec} that it is the collective mode which dominates the
spectrum. Thus, although the single particle features may be affected
by condensation, this would have little effect in practice. It may be
possible to isolate the single particle spectrum in a Rayleigh
scattering experiment, as has been proposed for equilibrium
condensates.\cite{marchetti_absorption_2007}

\section{Phase diagram}
\label{sec:phase-diagram}

\begin{figure}
  \centering
\includegraphics[width=8.6cm]{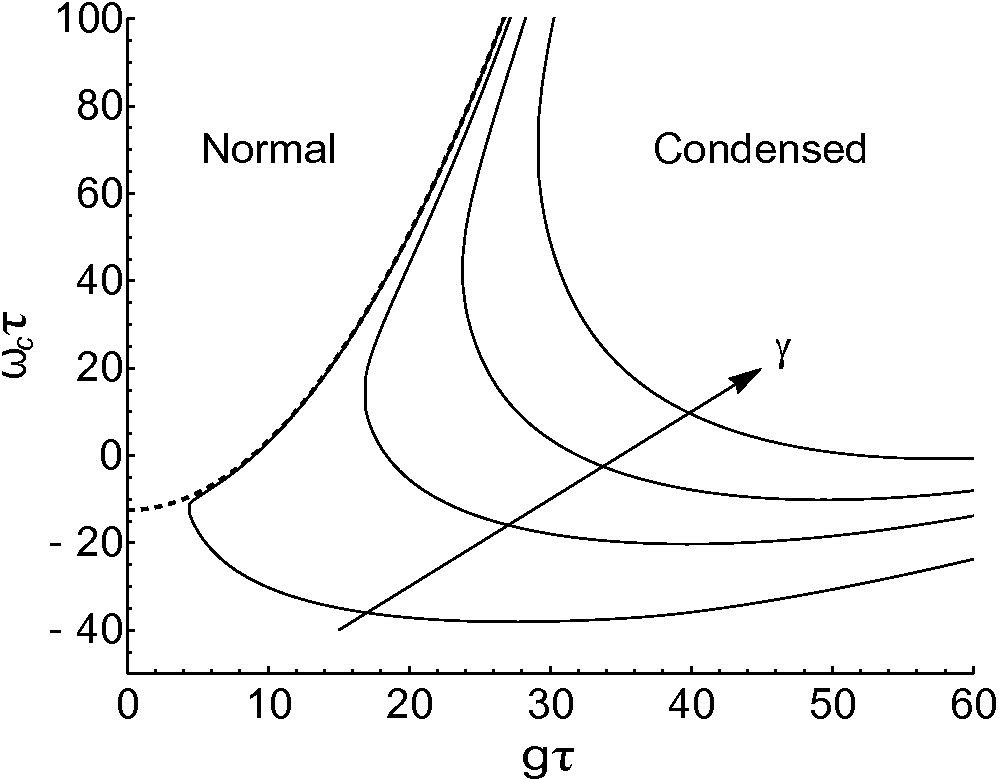}
\caption{Non-equilibrium phase diagram for a populated microcavity
  with a single photon mode as a function of coupling strength $g$ and
  cavity detuning $\omega_c$ with cavity damping $\gamma\tau=\text{1,
    15, 30, 45}$. Arrow indicates curves of increasing $\gamma$.
  Dotted line indicates the location of the equilibrium phase boundary,
  \cite{eastham_bose_2001} with temperature and chemical potential
  corresponding to that of the pumped population [$T=1/(k_B\pi\tau)$,
  $\mu=-12.5/\tau$].}
\label{fig:phase-diagram}
\end{figure}

\begin{figure}
  \includegraphics[width=8.6cm]{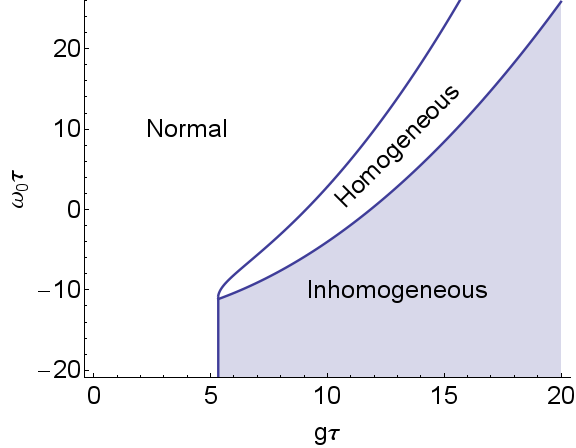}
  \caption{Non-equilibrium phase diagram for a
    populated microcavity with a continuum of in-plane photon modes as
    a function of the coupling constant $g$ and detuning at $\vec
    k=0$, $\omega_0$. The cavity damping $\gamma=1.5/\tau$. In the
    inhomogeneous region modes with $\vec k\neq 0$ have the highest
    growth exponent. The vertical boundary is a result of the
    gain-loss criterion, $H(\omega)>0$}
  \label{fig:phase-diagram-w0}
\end{figure}

The normal modes of the system, with frequencies determined by
$G(\omega)=0$, have decay rates $H(\omega)$. If a normal mode
frequency lies in the $H(\omega)<0$ region produced by the
non-equilibrium population it will be unstable, growing exponentially
to give a state with a highly populated mode, i.e., a condensate. The
condition for the onset of such an instability gives a non-equilibrium
phase diagram, which is shown for our chosen parameters in Figs.\
\ref{fig:phase-diagram} and \ \ref{fig:phase-diagram-w0}.

Fig.\ \ref{fig:phase-diagram} shows the phase diagram assuming that
only a single cavity mode, of energy $\omega_c$, is relevant. The
dotted line shows the phase boundary for equilibrium condensation in
the same model,\cite{eastham_bose_2001} with a temperature and
chemical potential corresponding to the pumped population. We see that
one sheet of the non-equilibrium phase boundary extends the
equilibrium result to allow for the cavity damping. Whereas in
equilibrium the presence of the collective mode is sufficient to
create an instability, in the open system condensation only occurs if
the gain at the energy of the collective mode overcomes the cavity
loss. Thus the collective mode can exist even in the normal state (see
Fig.\ \ref{fig:absspec}), and the damping pushes the transition to
larger couplings. In addition, we see that there is a lower limit on
$\omega_c$ in Fig. \ref{fig:phase-diagram}. This lower threshold is a
purely dynamical effect, not present in the equilibrium case. Below it
there is a bosonic collective mode at an energy well below that of the
populated states (see Fig. \ref{fig:absspec}). Although this mode
would be occupied in equilibrium it is far out of resonance with the
excitons. As a result, it is not occupied dynamically, and the
uncondensed state is metastable. A similar metastable region has been
predicted in quenched atomic gases.\cite{barankov_atom-molecule_2004}

\begin{figure}
  \centering
  \includegraphics[width=8.6cm]{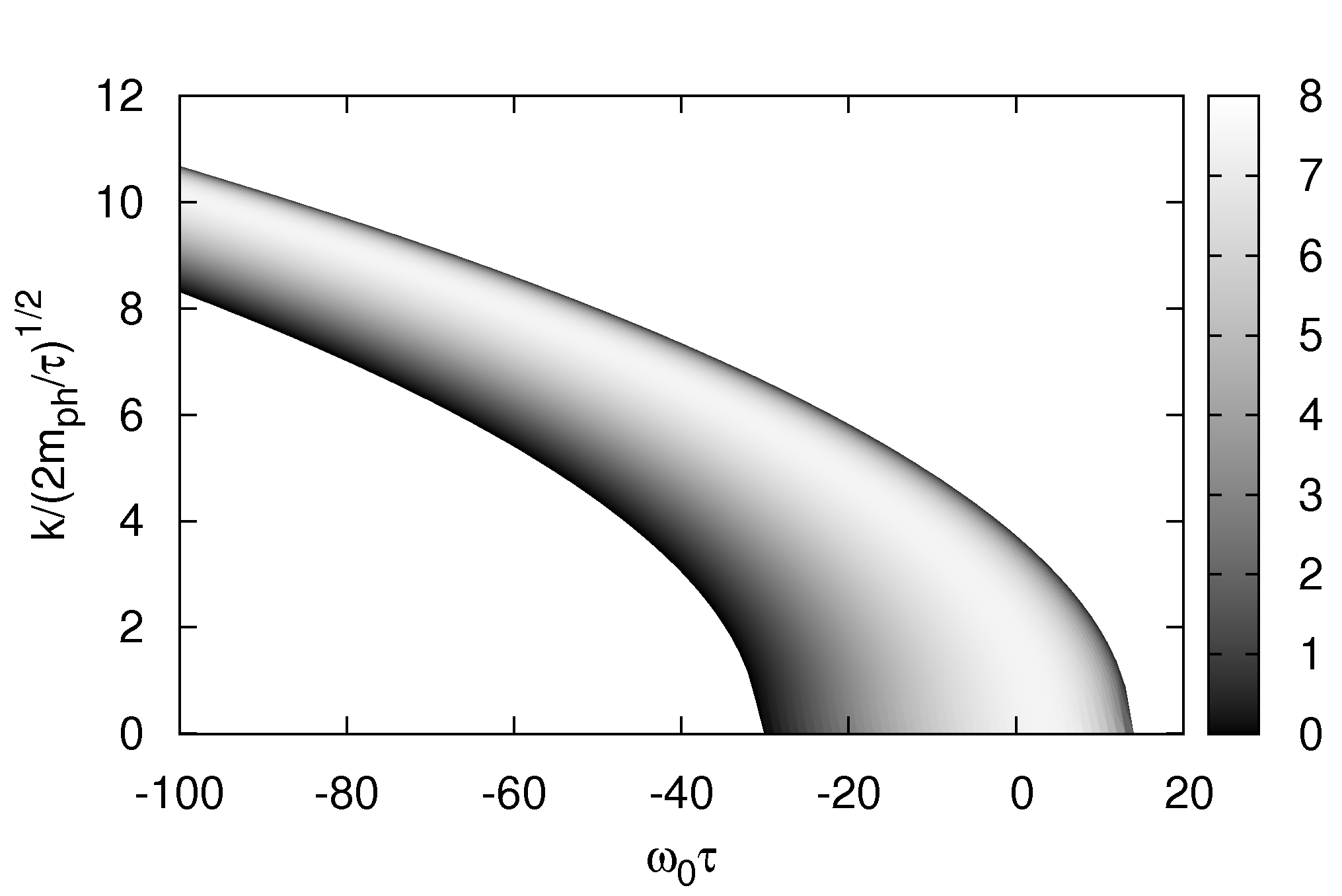}
  \caption{Growth exponents $-H(\omega)$ of the unstable modes with
    wavevector modulus $k$ in a populated cavity with detuning
    $\omega_0$, coupling strength $g=13/\tau$, and decay constant
    $\gamma=1.5/\tau$. The wavevector of the most unstable mode
    increases as $\omega_0$ is reduced. For $\omega_c\lesssim-30/\tau$
    the $\vec k=0$ mode is stable and there are only finite momentum
    instabilities.}
  \label{fig:growth-expon-unst}
\end{figure}

Figs.  \ref{fig:absspec} and \ref{fig:phase-diagram} show that, for a
given $g$ and $\gamma$, the condensation instability occurs over a
range of $\omega_c$. In a microcavity different values of
$\omega_c=\omega_0+|\vec k|^2/2m$ correspond to either changing
the cavity width, which varies the detuning $\omega_0$, or considering
modes at a different wavevector $\vec k$. As such, a range of unstable
$\omega_c$ implies that for a fixed cavity detuning there can be
instabilities at many wavevectors, with different growth exponents
$|H(\omega)|$. At short times after the population has been created
the mode with the highest growth exponent will dominate. Figure
\ref{fig:growth-expon-unst} shows that as $\omega_0$ is lowered this
dominant mode occurs at $\vec k\neq0$, implying a condensate with
finite momentum, and a spatially inhomogeneous order parameter. Thus
the full phase diagram, allowing for the continuum of in-plane modes,
takes the form shown in Fig.\ \ref{fig:phase-diagram-w0}.

The phase diagram of Fig.\ \ref{fig:phase-diagram-w0} can be
understood physically by noting that the condensation is a result of
the exciton-photon interactions. If the cavity mode is detuned well
below the excitons then the quasiparticles at $\vec{k}=0$ are
essentially photons, uncoupled from the excitons. Thus the
condensation shifts to the higher momentum states, where the photons
and excitons are closer to resonance, and there are strong coupling
effects.

\section{Discussion}
\label{sec:discussion}

There has been extensive theoretical work on states with finite
momentum Cooper pairing in the context of equilibrium superconductors,
atomic gases, and quark
matter.\cite{casalbuoni_inhomogeneous_2004} These FFLO states, which
have been sought in a wide range of systems, may be the ground states
where there is an imbalance in the populations of the two pairing
species. However, they involve increasing the kinetic energy in order
to gain pairing energy, and in practice this restricts them to small
regions of parameter space. Here, however, the state achieved is
determined by the Cooper equation and a gain-loss criterion, with the
energetics playing a subsidiary role. Thus, as indicated by
Fig. \ref{fig:phase-diagram-w0}, condensation at a finite momentum
may be achieved without fine tuning of parameters.

The connection to FFLO may be made more explicit using a
representation for the exciton operator $\hat\sigma_i^-=\hat
c_{i,\uparrow} \hat c_{i,\downarrow}$ where $\hat c$ is a fermionic
annihilation operator. In the simplest case of a plane wave condensate
at wavevector $\vec k$ the mean-field order parameter is $P_{\vec k}$,
the macroscopic component of the exciton polarization. This
becomes \begin{eqnarray} P_{\vec k}&=&(1/N) \sum_{i} \langle \hat
  c_{i,\uparrow} \hat c_{i,\downarrow}\rangle e^{-i \vec{k}.\vec{r}_i}
  \nonumber \\ &=& 1/N^2 \sum_{\vec{p},\vec{q}} \langle \hat
  c_{\vec{p},\uparrow} \hat c_{\vec{q},\downarrow}\rangle \sum_i e^{i
    (\vec{p}+\vec{q}).\vec{r}_i-i\vec{k}.\vec{r}_i} \nonumber \\
  &\sim& (1/N) \sum_{\vec p} \langle \hat c_{\vec{p},\uparrow} \hat
  c_{-\vec{p}-\vec{k},\downarrow}
  \rangle. \label{eq:maptofflo} \end{eqnarray} Thus we see that the
condensate can be formally represented as a coherent state of
fermions, pairing with a finite total momentum. While in this respect
the state is similar to FFLO, there are other differences. For
example, in Eq. (\ref{eq:maptofflo}) the relative wavefunction of the pair
is independent of momentum, and the pairing is entirely local. In a
general FFLO state there is a momentum dependent pairing function,
describing Cooper pairs of finite size.

Because the growth exponent depends only on $|\vec k|$ the condensate
emission at short times will cover a circle of in-plane wavevectors,
giving a cone of emitted light. However at later times the nonlinear
terms neglected in Eq. (\ref{eq:responpsistep}) will break the degeneracy,
selecting a spatial form for the condensate. In equilibrium such
interactions favor condensate structures consisting of a pair of
antipodal wavevectors $(\vec k,-\vec k)$, or more complex structures
such as face-centered cubes.\cite{casalbuoni_inhomogeneous_2004} Here
the nonlinearity corresponds to the depletion of the exciton
population by the growth of the condensate. This will reduce the
gain\cite{eastham_mode_2008} for collective modes of similar energies,
suggesting that a single plane wave (Fulde-Ferrel) state may be
favored. Although these nonlinearities determine a particular form
for the condensate it is unlikely they will lead to a homogeneous
state, so we do not treat them in detail here.

It is interesting to note that finite-momentum polariton condensates
have been observed,
\cite{richard_spontaneous_2005-1,lai_coherent_2007,krizhanovskii_coexisting_2009} though in a
different experimental protocol to that considered here. In these
cases there is continuous pumping and relaxation, and a spatial
structure imposed by a pump and trap. The mechanisms leading to this
finite momentum condensate have yet to be established, and are likely
different from those discussed here.  Nonetheless, these experiments
demonstrate that microcavities could support exotic ordered states
that have proved elusive in equilibrium.





\section{Summary}
\label{sec:conclusion}

We have calculated the linear response of a microcavity with a
non-equilibrium population of excitons. The population produces new
collective modes, which are analogs of the Cooper pairing mode in
superconductors. We have shown that these modes are visible as peaks
in the optical spectra. By considering the growth exponents of these
collective modes we have found a phase diagram for the dynamical
condensation. In a microcavity with a continuum of in-plane
wavevectors there can be multiple unstable modes of different
wavevectors. For some parameters the dominant (and, for sufficiently
negative detuning, only) instabilities can occur at a non-zero
wavevector. In these regimes the microcavity will develop a condensate
with spatial structure, signaled by coherent emission at an angle to
the cavity normal.

\begin{acknowledgments}
  This work was supported by Science Foundation Ireland Grant
  No. 09/SIRG/I1592 and EPSRC Grant No. EP/F040075/1. We thank
  P. B. Littlewood, J. Keeling, and M. Parish for discussions and
  comments on the manuscript.
\end{acknowledgments}

\appendix

\section{Analytical pump solution}
\label{sec:analyt-pump-solut}

Ref.~\onlinecite{eastham_quantum_2007} gives the results of numerical
simulations of Eqs.~(\ref{eq:eofm1}--\ref{eq:eofm3}), driven by a
linearly-chirped Gaussian pump pulse. These simulations show that
there are parameter regimes in which the dynamics separates into a
fast pumping stage, followed by a slower condensation stage. In this
appendix we present an approximate analytical solution to
Eqs.~(\ref{eq:eofm1}--\ref{eq:eofm3}) which gives the population and
(negligible) polarization at the end of the pump pulse. This solution
forms the starting point for the dynamics discussed in the body of the
paper.

The full numerical solutions in Ref.~\onlinecite{eastham_quantum_2007}
show that the only significant polarization during the pumping is at
the pump wavevector $\vec{p}$. Moreover, this polarization can be seen
to be small compared with the applied pump field $F$. Thus during the
pumping we may neglect the second term in Eq.~(\ref{eq:eofm1}) for all
wavevectors. With this approximation,
Eqs.~(\ref{eq:eofm1}--\ref{eq:eofm3}) reduce to an ensemble of
independent two-level systems, driven by a field $\psi^0_{\vec{p}}$
which is the externally applied field $F$ filtered by the cavity
response. For pumping at high angles, outside the stop band of the
mirrors, $\psi^0_{\vec{p}}$ is proportional to the pump pulse. Thus
Eqs.~(\ref{eq:eofm1}-\ref{eq:eofm3}) become the Bloch
equations\begin{equation}
\label{eq:bloch}
  \begin{pmatrix}
    \dot P'_x\\\dot P'_y\\\dot D_0
  \end{pmatrix}=
  \begin{pmatrix}
    0&E-\Delta(t)&0\\
    -E+\Delta(t)&0&g\Omega(t)\\
    0&-g\Omega(t)&0
  \end{pmatrix}
  \begin{pmatrix}
    P'_x\\P'_y\\D_0
  \end{pmatrix}, 
\end{equation}where we have defined \begin{gather} \psi^0_{\vec{p}}=\Omega(t)e^{-i\int^t\Delta(t)\,dt'},\\
  P'=P_{\vec p}e^{i\int^t\Delta(t)\,dt'},\\
  P'=\frac12(P_x-iP_y).
\end{gather} Note that the collective polarizations are at the pump
wavevector, while the collective inversion is spatially
uniform. 



%



For a model pump pulse of the form
\begin{gather}
\label{eq:pulse-params}
  g\Omega(t) = \frac {\Omega_0}{\tau}\sech\frac{t-t_0}{\tau}\text{,} \\
  \Delta(t) = \frac\alpha\tau \tanh\frac{t-t_0}{\tau}+\nu_0\text{,}\nonumber
\end{gather} Eq.~(\ref{eq:bloch}) has an analytical solution
\cite{hioe_two-state_1985}. The form of the population $D_0$ at times
$t\gg\tau$ after the pulse is:
\begin{widetext}
 \begin{equation}
  \label{eq:popdist}
  \frac{D_0(E)}{\nu(E)}={}2\frac{\cosh^2\frac{\pi\alpha}2
    -\cos^2\frac{\pi\sqrt{\Omega_0^2-\alpha^2}}2}
  {\cosh\left[\frac\pi2\big((E-\nu_0)\tau-\alpha\big)\right]
    \cosh\left[\frac\pi2\big((E-\nu_0)\tau+\alpha\big)\right]}-1\text{.}
\end{equation}
\end{widetext} In the limit $\Omega_0>\alpha\gg1$ the
distribution becomes
\begin{equation}
  \label{eq:popdistfermi}
  \frac{D_0(E)}{\nu(E)}=2n_F(E-\mu_+-\nu_0)\left(1-n_F(E-\mu_--\nu_0)\right)-1\text{,}  
\end{equation}
where $n_F(E)$ is a Fermi distribution with temperature $1/(k_B
\pi\tau)$ and the chemical potentials
$\mu_\pm=\nu_0\pm\frac\alpha\tau$. If the density of states $\nu(E)$
is sufficiently small at energies below $\mu_-$ then this lower edge is
irrelevant. The occupation function $D_0(E)$ is then equivalent to an
equilibrium Fermi distribution with $\mu=\mu_+$. In this paper we
consider parameters where this applies, choosing $\nu_0=-30/\tau$,
$\Omega_0=18$ and $\alpha=17.5$.

Since the dynamics during the pumping, Eq.~(\ref{eq:bloch}), involves
only $P_{\vec{p}}$, the polarization at any other wavevector $P_{\vec
  k\neq \vec p}$ remains zero. For the polarization at the pump
wavevector, the analytical solution gives a window of energies $\sim
\tau$ in which there is a non-zero polarization after
pumping. However, in the absence of an external field and with
$\psi_{\vec p}\approx0$, as is the case after pumping, the subsequent
evolution of the polarization is free. As a result, the total
polarization $P_{\vec p}=\int P_{\vec p}(E) dE$ decays by free
induction decay, and so may be neglected after a time of order
$\tau$. The numerical work of Ref.\ \onlinecite{eastham_quantum_2007}
showed that for suitable parameters the preparation of this state,
including the free induction decay of the remnant polarization,
finishes before the dynamics discussed in the main body of this paper
takes place.

\bibliographystyle{apsrev}
\bibliography{collectivemodesbib}

\end{document}